\documentclass[%
 reprint,
superscriptaddress,
 amsmath,amssymb,
pre,
]{revtex4-1}

\usepackage{epsfig}
\usepackage{graphicx}
\usepackage{dcolumn}
\usepackage{bm}

\begin{document}
\title{Computing shortest paths in 2D and 3D memristive networks}



\author{Zhanyou Ye}
\email{zhanyou.ye09@imperial.ac.uk}
\affiliation{Centre for Bio-Inspired Technology, Department of Electrical and Electronic Engineering, Imperial College London,
London SW7 2AZ, United Kingdom}
\author{Shi Hong Marcus Wu}
\email{marcuswu@mit.edu}
\affiliation{Systems Engineering Advancement Research Initiative (SEAri), Massachusetts Institute of Technology, 77 Massachusetts Avenue, E38-576 Cambridge, MA 02139-4307}
\author{Themistoklis Prodromakis}
\email{t.prodromakis@soton.ac.uk}
\affiliation{Centre for Bio-Inspired Technology, Department of Electrical and Electronic Engineering, Imperial College London,
London SW7 2AZ, United Kingdom}
\affiliation{School of Electronics and Computer Science, University of Southampton, Southampton SO17 1BJ, United Kingdom}



\date{\today}


\pacs{02.70.-c, 73.63.-b}

\maketitle

\onecolumngrid
Global optimisation problems in networks often require shortest path length computations to determine the most efficient route. The simplest and most common problem with a shortest path solution is perhaps that of a traditional labyrinth or maze with a single entrance and exit. Many techniques and algorithms have been derived to solve mazes, which often tend to be computationally demanding, especially as the size of maze and number of paths increase. In addition, they are not suitable for performing multiple shortest path computations in mazes with multiple entrance and exit points. Mazes have been proposed to be solved using memristive networks and in this paper we extend the idea to show how networks of memristive elements can be utilised to solve multiple shortest paths in a single network. We also show simulations using memristive circuit elements that demonstrate shortest path computations in both 2D and 3D networks, which could have potential applications in various fields. \\\ 

\twocolumngrid
\section{\label{sec:level1}INTRODUCTION} 
\indent  Many combinatorial optimisation problems in graph theory \cite{RefWorks:171}, such as the Travelling Salesman Problem, involve deriving the shortest path within networks \cite{RefWorks:161}. Applications of such computations include optimising routing protocols \cite{RefWorks:165}, transportation models \cite{RefWorks:160} and recurrent neural networks \cite{RefWorks:90}. Perhaps the simplest shortest path problem is a traditional maze, where one has to determine the path to the exit of a labyrinth whilst only given the entrance point. However, when there are a larger number of pathways in a maze, this increases the number of solutions. Out of these possible pathways, finding the shortest or least-cost one may not necessarily be straightforward. Many mathematical algorithms have been proposed to solve mazes, such as random mouse or mathematical search algorithms \cite{RefWorks:162, RefWorks:93}. Such algorithms derive solutions in a sequential fashion, thus solution times can increase exponentially in complex networks. \par \indent
There are also many innovative methods prescribed to solve mazes using biological and chemical systems, such as amoeboid organisms \cite{RefWorks:4,RefWorks:3}, chemotaxis \cite{RefWorks:87} and chemotactic droplets \cite{RefWorks:7}. However, such methods also suffer from increased time complexity when maze sizes increase. The problem is further exacerbated with the introduction of multiple users to a network, such as a traffic optimisation problem where multiple cars would like to find the shortest travelling path in order to avoid congestion in the network. In this paper, we propose using networks of memristive elements to perform multiple shortest path computations in a given network. \par \indent
 The memristor, short for memory resistor, is a passive two-terminal circuit element capable of altering its resistance based on the input and remember its past dynamics \cite{RefWorks:5}. After the device was postulated by L.Chua, a generalised concept of the memristor was further proposed by Chua and Kang \cite{RefWorks:173}, defined as  
\begin{equation}
 v = R(x)i
\end{equation}
\begin{equation}
\frac{dx}{dt} = f(x,i)
\end{equation}
where \begin{math} v \end{math} represents the voltage, \begin{math} i \end{math} represents the current and  \begin{math} R(x) \end{math} denotes the instantaneous resistance of the device that changes based on its internal state variable,  \begin{math} x \end{math} \cite{RefWorks:174}. Memristance signatures are also observed in various dissipative systems that support discharge phenomena, such as discharge lamps and biological ion channels \cite{RefWorks:20, pershin2011memory}. Since its implementation by Strukov et.al \cite{RefWorks:22}, the solid-state memristor has been proposed to be of use in various applications such as memory storage \cite{RefWorks:163} and neuromorphic implementations \cite{RefWorks:159, RefWorks:158}. \par \indent
Memristor networks - several memristors connected in the form of an array- have been postulated to be able to perform complex cortical computing functions \cite{RefWorks:158}. Pershin and Di Ventra have also demonstrated that abstract mazes can be solved in a parallel fashion using memristive networks, a termed coined as \emph{analog parallelism}. They have also shown that all solutions in the maze can be determined and the results are separated in order of path length \cite{RefWorks:1}. \par \indent
In the rest of the paper, we exploit the plasticity of 2D and 3D memristive networks for extrapolating various shortest path solutions via simulations in PSPICE. Our study initiates by deciphering the fundamentals of the network to derive shortest path solutions to a given maze in 2D. We then expand this concept to exhibit how a memristive grid can be used to perform multiple shortest path computations in a network involving several users; the example here is London's Tube Network. Lastly, we show how multiple shortest path problems can be solved using 3D memristive networks. 
\section{METHODOLOGY} 
\indent The maze or network must first be mapped onto a regular memristive grid. The representative memristive grid is then implemented in MATLAB and the corresponding circuit simulations are performed using PSPICE. Entrance/Exit or Start/End nodes for the circuit simulation are represented by a 1V DC Voltage source and Ground respectively. 
\subsection{MEMRISTIVE COMPONENTS} 
\indent In 1971, the memristor was predicted theoretically by L. Chua in his seminal paper \cite{RefWorks:5} but it remained a theoretical abstraction until researchers at Hewlett Packard (HP) Laboratories discovered similar properties while fabricating crossbar-type nano-devices in 2008 \cite{RefWorks:22, RefWorks:92}. The memristor was postulated based on a mathematical relationship between charge \begin{math}q\end{math} and magnetic flux \begin{math} \varphi \end{math} : \begin{math}d\varphi = Mdq\end{math}, where \begin{math}M\end{math} denotes the memristance, which has the same units as resistance (\begin{math} \Omega \end{math}) and is defined as the resistance across the memristor. By taking the time integrals of \begin{math}q\end{math} and  \begin{math} \varphi \end{math}, the non-linear relationship between voltage and current across the memristor is established:
\begin{equation}
 v(t) = M(q) *  i(t)
\end{equation}
In Eq (3), \begin{math}v(t)\end{math} is the applied bias, \begin{math}i(t)\end{math} is the current flowing through the memristor and \begin{math}M(q)\end{math} is the charge-dependent memristance. The simplest abstraction of the memristor is that of a time-dependent resistor \cite{RefWorks:64}:
\begin{equation} M(t) =  \frac{W(t)}{D}*R_{ON} + (1 - \frac{W(t)}{D})*R_{OFF} \end{equation}
\subsection{MEMRISTIVE FUSE}
The conductance modulation of a single memristor depends on the polarity of the charge flowing through it. However, when devices are used in memristive networks for shortest path computations, polarity dependence is not desirable since the direction of current flow cannot always be determined. \par 
\indent It was previously proposed that by connecting two memristors with opposing polarities \cite{RefWorks:172}, the non-linear relationships between time integrals of voltage and current can be preserved without any polarity dependence. This new combination of devices is termed the memrsitive fuse \cite{RefWorks:17}, shown in Fig. 1, and is used as the primary memristive device in all the networks in the following simulations. Since the fuse is made of two ideal memristors connected in series, the total initial resistance of the overall device is twice the intial resistance of a single memristor (\begin{math} 2R_{INIT} \end{math}).
\begin{figure}
\centerline{\includegraphics[scale=0.6]{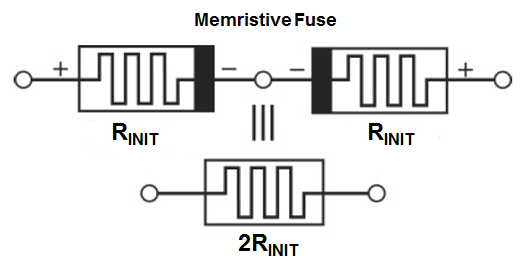}}
\caption{Two memristors connected with opposite polarities to form a Memristive Fuse \protect\cite{RefWorks:17}.}
\end{figure}
\subsection{SPICE SIMULATIONS}
All the memristive networks are simulated using Biolek's Memristor SPICE Model \cite{RefWorks:21} with Prodromakis' non-linear kinetics dopant model \cite{RefWorks:18}. Details about the models can be found in the corresponding references. Throughout this study, the memristor parameters within the SPICE model were defined as follows: Initial Width \begin{math} W_{0}  = 5 \times 10^{-9} \end{math}m, Active-Core Thickness \begin{math} D  = 10 \times 10^{-9} \end{math}m, ON Resistance  \begin{math} R_{ON} =100 \space \Omega \end{math}, OFF Resistance  \begin{math} R_{OFF}  = 16  k\Omega \end{math}, Net Resistance at t = 0 \begin{math} R_{INIT}  = 1000  \Omega \end{math}, Mobility  \begin{math} \mu  = 1 \times 10^{-14} m^{2}s^{-1}V^{-1}\end{math}.\\
\section{SHORTEST PATH SOLUTION OF MAZES USING 2D MEMRISTIVE NETWORKS} 
The first simulation shows the varying conductance paths in a memristive network, which correspond to the various solutions to a simple shortest path computation. A simple memristive network is first constructed from a combination of memristive fuses and resistors. Throughout this work we refer to the points where devices are connected together as nodes, while we refer to a branch in the case it comprises one or more devices between two nodes. In Fig. 2, a simple maze is illustrated using a 4 x 4 memristive network. The paths of the maze are simulated using 12 memristive fuses (labelled M1 - M12) and 2 \begin{math} M\Omega \end{math} resistors are used to represent the blocked conductance paths. A 1V DC Voltage Source and Ground are placed at the nodes corresponding to the entrance and exit of the maze respectively. \par \indent
\begin{figure}
\centerline{\includegraphics[scale=0.6]{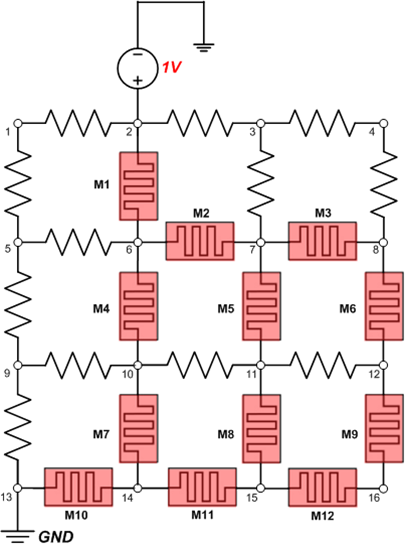}}
\caption{(Colour Online) 4 x 4 Memristive Network consisting of Memristive Fuses and Resistors.}
\end{figure} 
The memristive network is simulated for 35s and the results are shown in Fig. 3 and Fig. 4. For each device, the change in memristance \begin{math} \varDelta M \end{math} is determined by taking the difference between the ‘resistance’ across each branch and the initial resistance (\begin{math}2R_{INIT}\end{math}), which is calculated via:
\begin{equation} \varDelta M_k = \vert \frac{V_X - V_Y}{I_{XY}} \vert - 2R_{INIT} \end{equation}
\begin{math}x\end{math} and \begin{math}y\end{math} are the nodes connecting each branch and \begin{math}k\end{math} is the device number. Fig. 3 illustrates the transient response of   \begin{math} \varDelta M \end{math} for devices M4, M5 and M6 against time. This change shows how the devices in different paths respond to the input voltage due to the variance in current amplitudes flowing through them. Fig. 4 shows the temporal evolution of memristance of all 12 devices in the network for time instances 1s, 5s, 10s and 30s. For better visualisation of the change in memristance,  \begin{math} \varDelta M \end{math} for each device was translated to a linear colour scale of 0 - 64, where 0 corresponds to zero \begin{math} \varDelta M \end{math} and 64 represents the maximum  \begin{math} \varDelta M \end{math} observed throughout the duration of the simulation. \par \indent
We first analyse the memristance change of the devices between three branches: Branch 1 (nodes 6 and 14), Branch 2 (nodes 7 and 15) and Branch 3 (nodes 8 and 16). Kirchoff’s Current Law states: 
\begin{equation} I_{in} = \sum^{n}_{i=1} I_i\end{equation}
\begin{math}n\end{math} refers to the number of branches at the particular node. Applying the formula at nodes 14 and 15, the following relationship regarding the overall current flow across all three branches can be deduced:  \begin{math}I_{B3}\end{math} \textless \begin{math}I_{B2}\end{math} \textless \begin{math}I_{B1} \end{math}. \par \indent
\begin{figure}
\centerline{\includegraphics[scale=0.3]{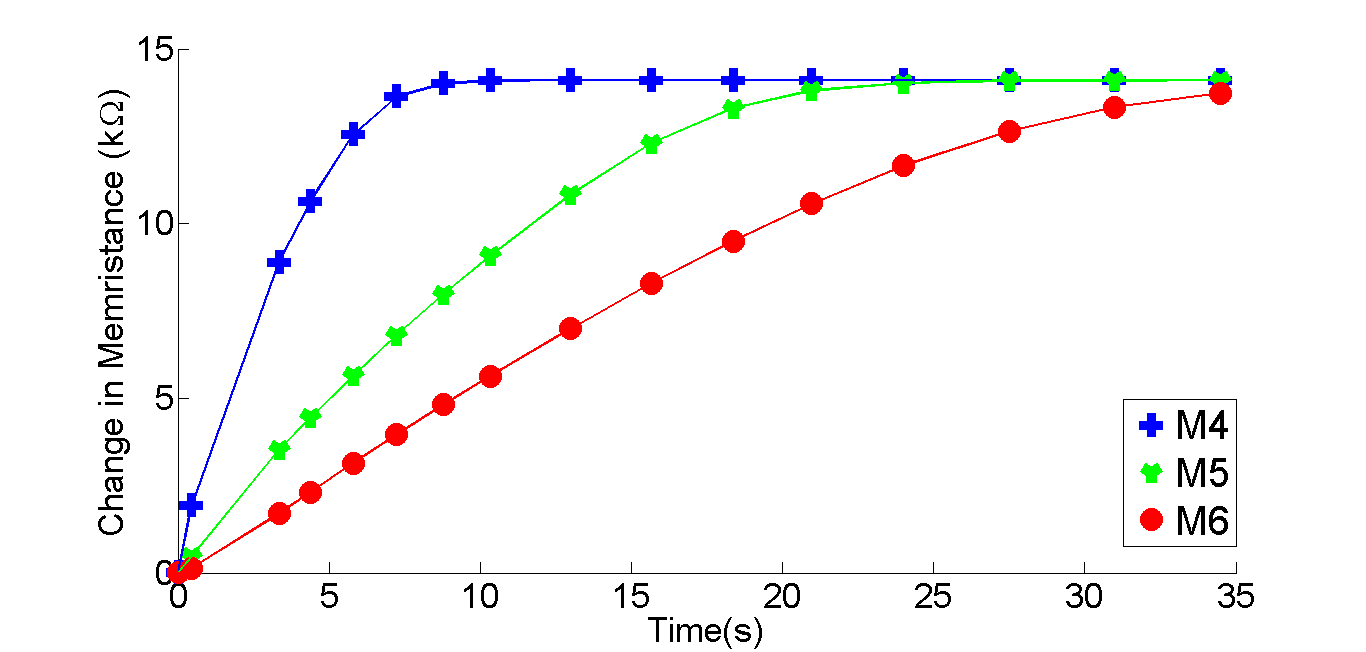}}
\caption{(Colour Online) Plot of Memristance Change (Devices M4, M5 and M6) for time period 0 - 35s.}
\end{figure}
A larger current flow across a memristor will result in higher rate of change of memristance of the device. Assuming that very little current flows through the 2 \begin{math} M\Omega \end{math} resistors in the grid network, this implies that the \begin{math} \varDelta M \end{math} of the devices in all three branches after a short time period will have a similar relationship to that of the total initial current flow across the branches:  refer to the change in memristance across devices M4, M5 and M6, shown in Fig. 3 respectively. An increase in the memristance across the devices in Branch 1 will in turn channel more current to Branches 2 and 3. After a stipulated simulation time, all the memristive devices in the network will reach the high resistive state (\begin{math}R_{OFF}\end{math}). \par \indent
\begin{figure*}
\includegraphics[scale=0.8]{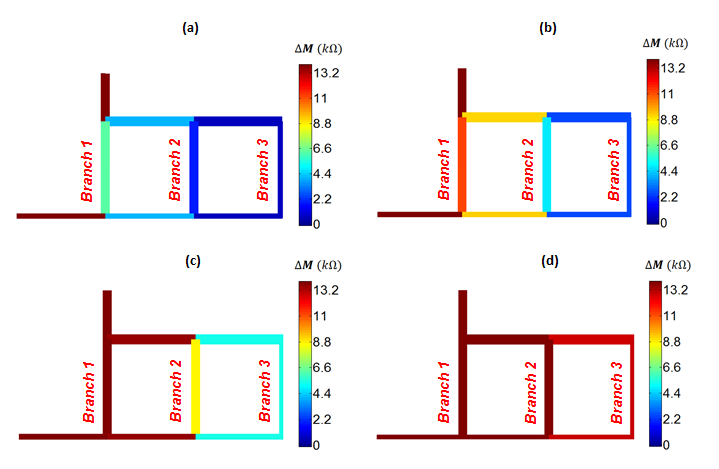}
\caption{(Colour Online) Spatiotemporal plot depicting \begin{math} \varDelta M \end{math} of all Memristive Elements at varying times: (a) 1s, (b) 5s, (c) 10s, (d) 30s. The colour bar on the right show the corresponding \begin{math} \varDelta M \end{math} values.}
\end{figure*}
\begin{figure*}
\centerline{\includegraphics[scale=0.8]{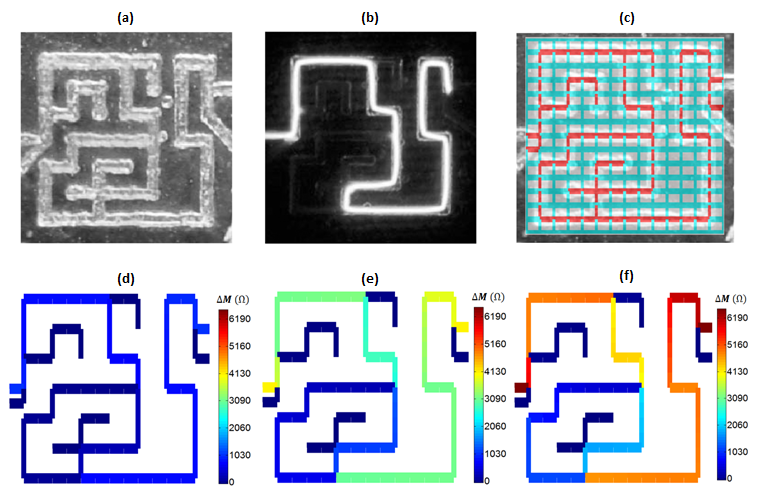}}
\caption{(Colour Online) Maze (a) and Solution (b) shown by Reyes et al \protect\cite{RefWorks:15}. (c) Mapping of Maze to a 15 x 15 Memristive Grid. Solutions to maze shown by simulations after 2s (d), 6s (e) and 10s (f).}
\end{figure*}
\begin{figure*}
\centerline{\includegraphics[scale=0.72]{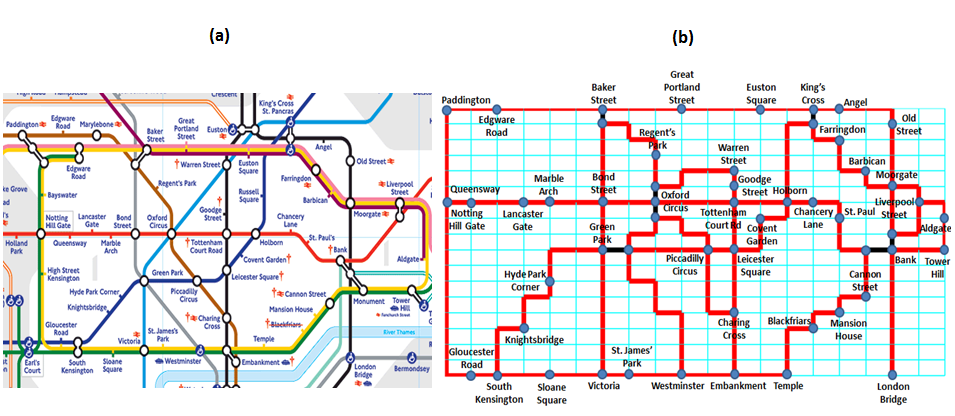}}
\caption{(Colour Online) Zone 1 of London's Tube Network (a) is mapped onto an 18 x 20 Memristive Grid (b).}
\end{figure*}
It can be observed that the shortest path in a maze will exhibit a larger change in memristance over a single period. For the simple maze in Fig. 2, Branch 1 is clearly the shortest path, and the corresponding length of the other two paths (Branch 2 followed by 3) can be identified by comparing  \begin{math} \varDelta M \end{math} of all devices. We also note the limit of the simulation where all devices reach the high resistive state (\begin{math}R_{OFF}\end{math}) and the paths are no longer distinguishable by measuring  \begin{math} \varDelta M \end{math}. Based on the argument that discharge-phenomena support memristive signatures \cite{RefWorks:20}, we have reviewed numerous reports on unconventional computation via discharge mechanisms; the most prominent one being \cite{RefWorks:15}. The maze, shown in Fig. 5a, is reproduced by Reyes et.al \cite{RefWorks:15}, where the solution is determined using an analog computation method via glow discharge in microfluidic chips (Fig. 5b). The maze is first mapped onto a 15 x 15 memristive grid and the red and blue lines of the grid overlapped onto the maze represent memristive fuses and 2 M\begin{math} \Omega \end{math} resistors respectively, as shown in Fig. 5c. A 1V DC Voltage Source and Ground are placed at the entrance and exit nodes respectively. Spatiotemporal representation of the change in memristance \begin{math} \varDelta M \end{math} of the memristive devices are shown in Fig. 5d, Fig. 5e and Fig. 5f for the times 2s, 6s and 10s respectively, and the shortest path solution is shown to be identical when compared with the solution derived using microfluidic chips.  \par \indent
This case verifies that the solution to a maze can indeed be determined by mapping it to a memristive grid and placing the source and ground at the entrance/exit nodes.  At the same time, this example proves the concomitantly argument presented in \cite{RefWorks:20}: discharge phenomena manifest memristive signatures. By exploiting the analog computations facilitated by Kirchoff's Current Law and that current follows the shortest path to ground, the shortest conductance path will exhibit the largest \begin{math} \varDelta M \end{math}. In addition, the altered devices will stay at their given resistive states even after the source and ground nodes have been removed. \par \indent
In this example, the memristive network converged to a possible solution to the maze after a simulation time of approximately 6 - 10s. Nonetheless, this approach is clearly amenable to the use of larger biasing potentials that will in turn speed up the solution. It is interesting to compare this to other mathematical search algorithms performed by a mirco-mouse robot; a robot searching for the shortest path in a 16 x 16 unit square maze using either Dijkstra's \cite{ RefWorks:180} or Flood-Fill algorithms typically requires 100s of seconds to accomplish similar tasks \cite{RefWorks:93,RefWorks:89}. Although actual memristive hardware implementations may yield different solution times from software simulation results, this comparison gives us a scale of the improvements in time complexity by utilising such memristive networks. 
\begin{figure*}
\centerline{\includegraphics[scale=0.7]{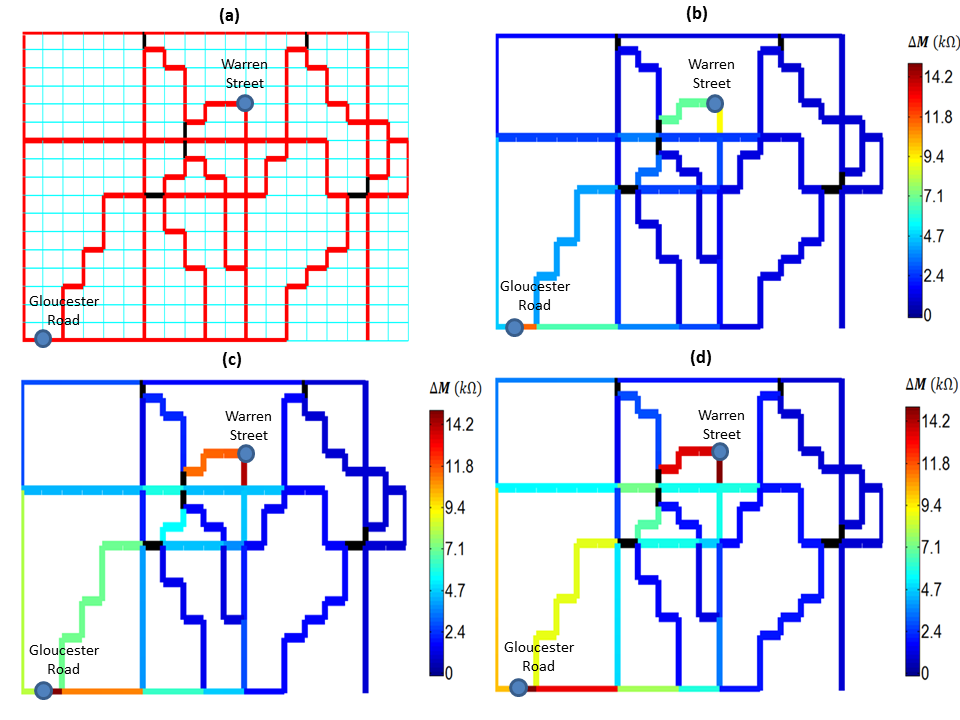}}
\caption{(Colour Online) Traveller A heading from Gloucester Road to Warren Street Station. The respective positions of nodes are shown on the network in (a). Solution to the network shown by simulations after 3s (b), 7s (c) and 10s (d). }
\end{figure*}
\section{MULTIPLE SHORTEST PATH COMPUTATIONS USING 2D MEMRISTIVE NETWORKS}
\indent So far, we have seen the computation of shortest paths for mazes with fixed entrance and exit points. In this section, we further elaborate on the possibility of concurrently solving multiple shortest paths within a same network via an example of travellers determining the shortest path on London's Tube Network. Zone 1 of London's Tube Map, shown in Fig. 6a, is first mapped onto an 18 x 20 memristive grid, as illustrated in Fig. 6b. Similarly, the red and blue lines on the grid represent memristive fuses and 2 M\begin{math} \Omega \end{math} resistors respectively. The mapping is an approximation of the actual distances and time taken between the tube stations and solely for the demonstration of shortest path computations between stations. \par \indent
One of the limitations of performing simulations using 2D memristive grids is that each centre node and corner node can accommodate a maximum of four and two paths passing through them respectively. While investigating the Tube Network application, this limitation in paths per node is insufficient for representing some stations such as Green Park, which has six lines going in and out of the station. 
Hence, in order to increase the number of possible paths through each node without increasing the dimensional space of the network, a 1\begin{math} \Omega \end{math} resistor is hereby used to link two neighbouring nodes to increase the node size. These extensions are shown as black lines in Fig. 6b and they signify that the two nodes are now effectively the same station. In this scenario, the voltage drop across the resistor is assumed to be negligible since the corresponding resistance is three orders of magnitude smaller than the initial resistance \begin{math}R_{INIT}\end{math} of the memristors used in the circuit.  All starting and destination nodes in the memristive network are simulated using 1V DC Voltage Sources and Ground. \par \indent 
We first show the shortest path computation in the Tube network for a single traveller wishing to get from Gloucester Road Station to Warren Street Station. Fig. 7 shows the corresponding results of the network for the times 3s, 7s and 10s and the shortest path is accurately determined by observing the spatiotemporal plot of  \begin{math} \varDelta M \end{math} for all the memristive devices in the network. Moreover, we demonstrate how the memristive network computes shortest paths for three travellers, namely Travellers A, B and C in the tube network concurrently. In the first scenario, all three travellers wish to get to the same destination, Holborn from their respective starting stations: A - Paddington, B - Gloucester Road and C - London Bridge. At the circuit level implementation of the memristive network, this translates to three 1V DC sources at the starting nodes and a single ground placed at the node representing Holborn station. The shortest paths of the three travellers will be termed \begin{math} P_A\end{math}, \begin{math}P_B\end{math} and \begin{math}P_C\end{math} respectively and are shown in the memristive network in Fig. 8. \par \indent
\begin{figure*}
\centerline{\includegraphics[scale=0.7]{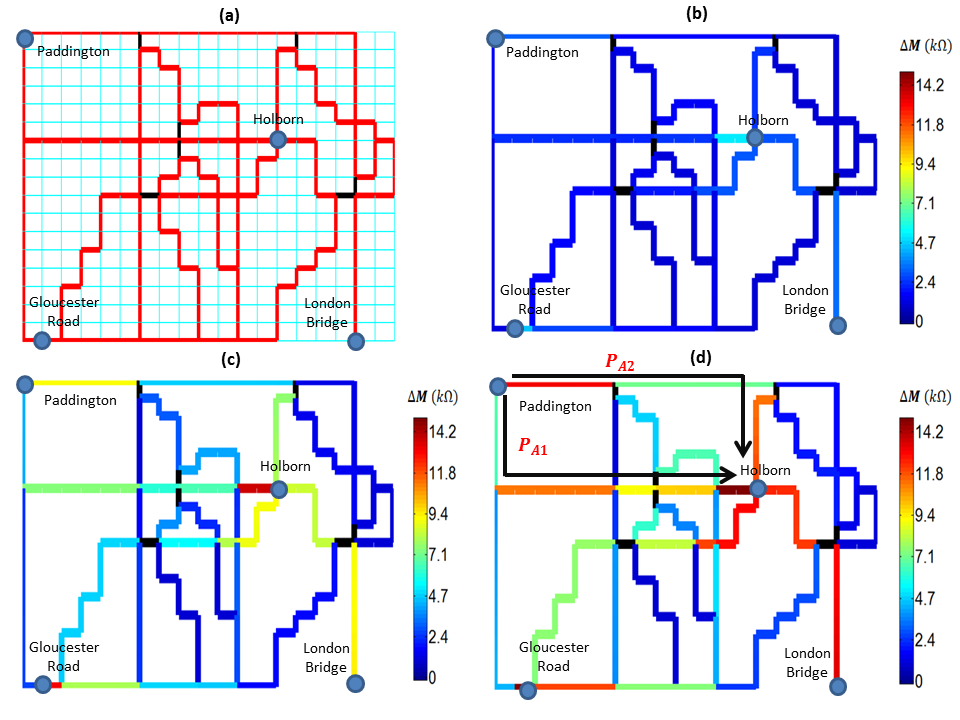}}
\caption{(Colour Online) Travellers A, B and C heading from Paddington, Gloucester Road and London Bridge Stations to Holborn Station. The respective positions of nodes are shown on the network in (a). Solution to the network shown by simulations after 1s (b), 5s (c) and 10s (d). Black arrows shown in (d) indicate the two paths for Traveller A (Paddington to Holborn).}
\end{figure*}
\begin{figure*}
\centerline{\includegraphics[scale=0.7]{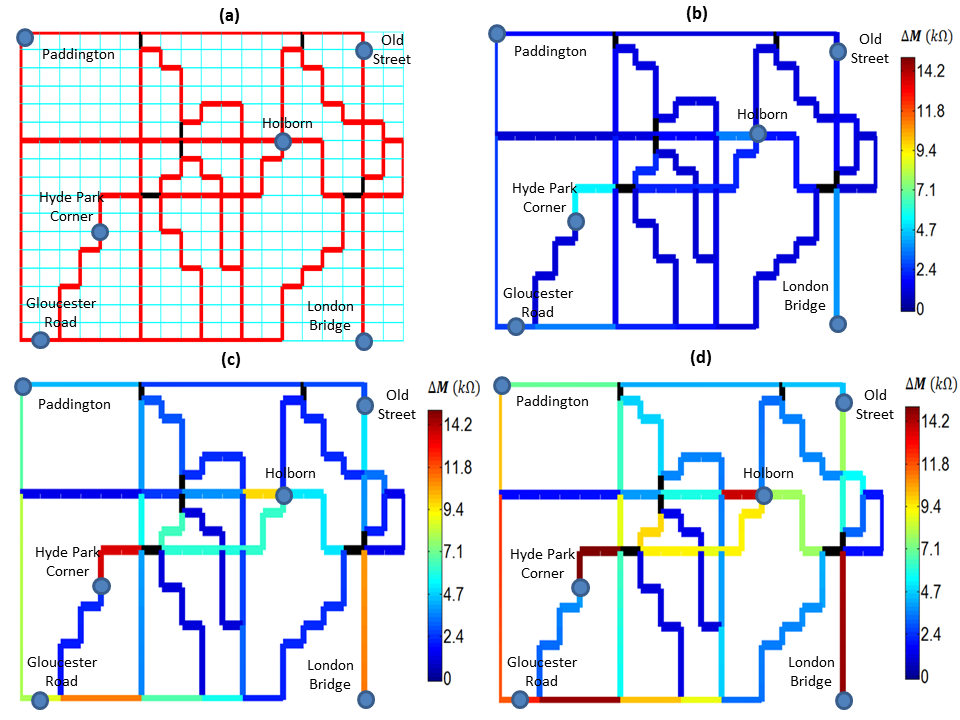}}
\caption{(Colour Online) Travellers A, B and C on three routes: Gloucester Road to Paddington, Hyde Park Corner to Holborn and London Bridge to Warren Street. The respective positions of nodes are shown on the network in (a). Solution to the network shown by simulations after 1s (b), 5s (c) and 10s (d).}
\end{figure*}
\begin{figure*}
\centerline{\includegraphics[scale=0.7]{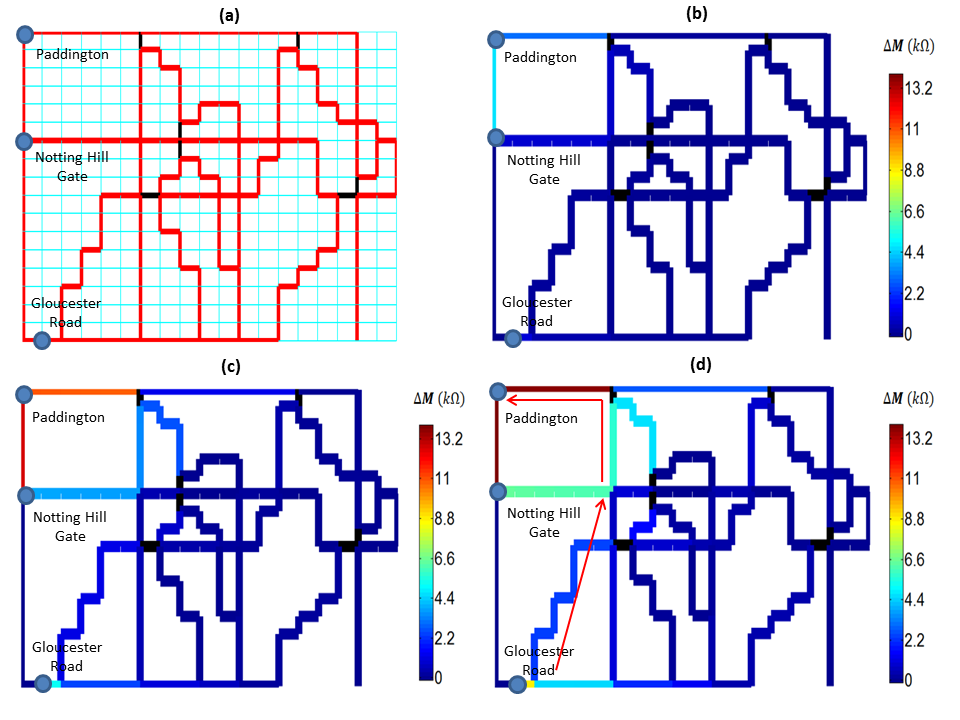}}
\caption{(Colour Online) Travellers A and B heading from Gloucester Road and Notting Hill Gate to Paddington Station. The respective positions of nodes are shown on the network in (a). Solution to the network shown by simulations after 1s (b), 5s (c) and 10s (d). The red arrow shown in (d) indicates the alternate path for Traveller A.}
\end{figure*}
By comparing the relative \begin{math} \varDelta M \end{math} on the spatiotemporal plot shown in Fig. 8, we note that there are two possible shortest paths solutions, \begin{math}P_{A1}\end{math}  and \begin{math}P_{A2}\end{math}. The number of memristive elements (\begin{math}N\end{math}) for the two solutions are \begin{math}N_{A1} = 19\end{math}  and \begin{math}N_{A2} = 20\end{math}, as shown in Fig. 8. As the memristive network size increases, the average \begin{math}N\end{math} increases as well. If the difference in path lengths, \begin{math}(N_1 - N_2 )  \ll N_x\end{math}   (where \begin{math}x\end{math} = 1 or 2), it will be increasingly difficult to distinguish between two shortest paths using \begin{math} \varDelta M \end{math} of the devices as \begin{math} \langle\varDelta M_{P1}\rangle \approx \langle\varDelta M_{P2}\rangle \end{math}, where \begin{math}\langle\varDelta M_{P1}\rangle\end{math} and \begin{math}\langle\varDelta M_{P2}\rangle\end{math} are the average \begin{math} \varDelta M \end{math} of the memrisitive devices in paths 1 and 2 respectively. In the second scenario however, shown in Fig. 9, Travellers A, B and C all have different start and end stations: Traveller A wishes to get from Gloucester Road to Paddington, Traveller B from Hyde Park Corner to Holborn and Traveller C from London Bridge to Old Street. It is noted that all computed paths are unique solutions; there are no overlapping paths between the travellers. \par \indent
Due to the use of voltage sources at the starting node of the route, the shortest path computed for one traveller will not pass through the starting point of another. This is shown in another example, when Travellers A and B travel from Gloucester Road and Notting Hill Gate to Paddington respectively. As seen from the shortest path computations presented in Fig. 10, the path computed by the memristive network for Traveller A does not pass through Notting Hill Gate station although that path has a lower \begin{math} N\end{math} value. In the circuit implementation, both station nodes are at high voltage potential, hence resulting in a negligible amount of current flow between them. Even after 10s, the measured \begin{math} \varDelta M \end{math} of the devices between the two nodes is approximately only 2\begin{math}\Omega\end{math}. The shortest path for Traveller A will essentially be the next alternative path, as marked out by the red arrow in Fig. 10. \par \indent
This series of cases exhibit that multiple shortest path computations can be performed based on the overall change in memristance due to the current flows in a single 2D memristive network. This has been demonstrated using London's Tube Network, where the shortest paths of three travellers are determined concurrently using a single network. If other known shortest path algorithms such as Dijkstra's \cite{ RefWorks:180} were used in this example, routes for the three travellers will have to be determined independently, which increases the time complexity of computation by an order of the number of travellers there are in a network. By computing the shortest paths in a parallel manner by solving a series of Kirchoff's Current Law equations, the memristive grid is able to compute all shortest paths in a single step. In addition, all the solutions are shown over a fixed time period regardless of the number of travellers in the network. \par \indent

\section{SHORTEST PATH COMPUTATIONS USING 3D MEMRISTIVE NETWORKS}
\indent The limitations using 2D networks are fewer input and output paths per node, in addition to the relatively low spatial resolution that can be achieved. For example, if all the lines in London's Tube Network (Zones 1-5) were to be mapped onto a single memristive network, it will be more accurately performed in 3D, where an additional layer can accommodate overlapping lines in the Tube network. In this section, we describe the computation of shortest paths by employing 3D grids, using a simple maze constructed in a 4 x 4 x 3 3D memristive network with two entrances and a single exit. The paths for the maze, represented using memristive fuses in the corresponding circuit are shown in Fig. 11a as light blue lines, while all static resistive elements are represented by thin black lines. The corresponding circuit is exploited in a similar manner to the pre-discussed scenarios. The two shortest path solutions of the maze, shown in Fig. 11b, Fig. 11c and Fig. 11d, are clearly depicted by monitoring the \begin{math} \varDelta M \end{math} of all memristive devices. \par \indent
\begin{figure*}
\centerline{\includegraphics[scale=0.7]{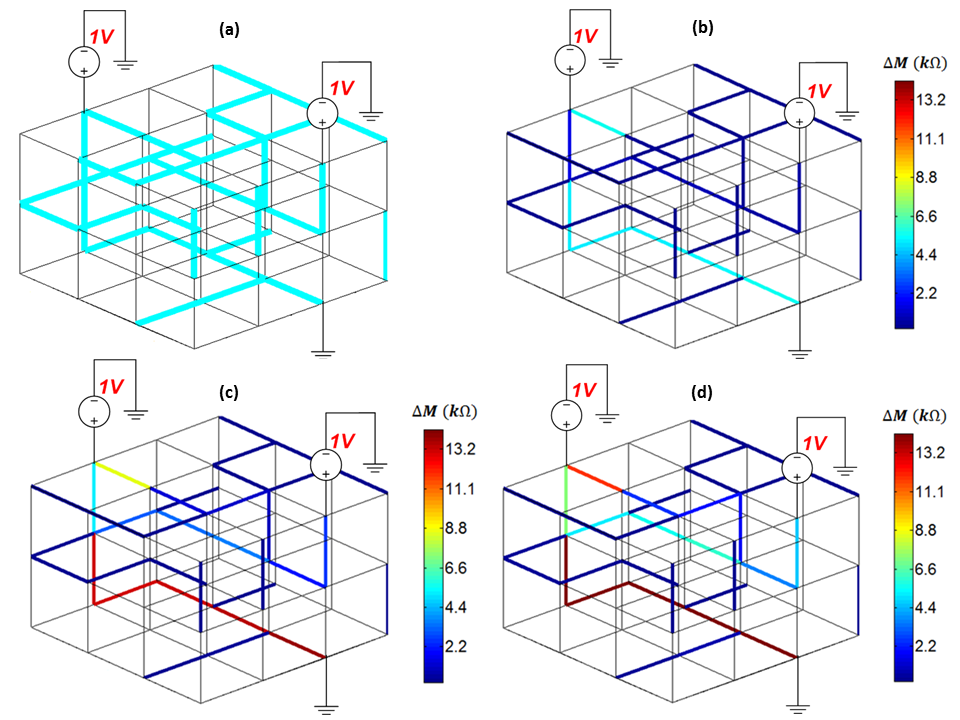}}
\caption{(Colour Online) Maze shown in a 4 x 4 x 3 Memristive Network. Paths of the maze highlighted in light blue (a), with entrances and exits indicated by 1V and Ground respectively. Solution to the network shown by simulations after 1s (b), 5s (c) and 10s (d). }
\end{figure*}
A 3D network can also be viewed as several 2D arrays stacked onto each other, with the addition of linking elements between the layers. We compare the time complexity of solving a 3D maze if any random mouse algorithm is used \cite{RefWorks:93}. Assuming that the number of vertices and paths in each layer remain the same, the total time complexity for the random mouse method will increase by an order of the number of 2D arrays, including the number of interconnecting paths. \par \indent
This computation method via 3D memristive networks has been proven to be simple to execute and does not require long computation times. Current shortest path algorithms such as Dijkstra's \cite{ RefWorks:180} and Floyd-Warshall's \cite{ RefWorks:181} have time complexities of \begin{math}O(V^2)\end{math} and \begin{math}O(V^3)\end{math} respectively where  \begin{math}V\end{math} is the number of vertices (nodes) \cite{RefWorks:84}. In comparison to the employed memristive network, the best theoretical estimate for a linear system is the Coppersmith Winograd algorithm \cite{RefWorks:91,RefWorks:1} which is described as \begin{math}O(n^{2.376})\end{math} where \begin{math}n\end{math} is the number of edges in a network. However, it is noted that a memristive network implemented in hardware only has a single overall computation step in order to determine the shortest paths in the network \cite{RefWorks:1}. This makes it more efficient than the algorithms listed above, before even considering multiple computations in a single network. \par
\section{CONCLUSION}
\indent Paths of an existing network can be mapped on a memristive network using a series of memristive devices and resistors. By exploiting the analog computations performed by solving Kirchoff's Current Laws in a parallel manner \cite{RefWorks:1}, memristive networks have been shown to be capable of computing shortest paths in a given maze, leveraging on the dynamic adjustment of their intrinsic conductance. This computation method has also been extended to show how multiple computations can be performed. Furthermore, this concurrent solution method can also be exploited to include 3D spaces, where shortest paths through stacks of 2D arrays can be efficiently determined by performing a single step via employing distinct voltage sources and ground terminals to the entrances and exits of the network. Such networks, if implemented in hardware, have great application prospects and can be used to solve many optimisation problems in various fields.

\begin{acknowledgments}
The authors wish to acknowledge the financial support of the CHIST-ERA ERAnet EPSRC EP/J00801X/1 and EP/K017829/1.
\end{acknowledgments}

\bibliographystyle{apsrev4-1}
\bibliography{shortest_path}

\end{document}